\documentclass[aps,prb,showpacs,superscriptaddress,twocolumn,citeautoscript,floatfix]{revtex4-1}
\usepackage{graphicx}
\usepackage{amsmath}
\usepackage{hyperref}

\begin{document}

\title{Speed limit to the Abrikosov lattice in mesoscopic superconductors}

\author{G.~Grimaldi}
\email[]{gaia.grimaldi@spin.cnr.it}
\affiliation{CNR SPIN Salerno, via Giovanni Paolo II, 132, 84084 Fisciano (SA), Italy}
\author{A.~Leo}
\affiliation{Physics Department ``E. R. Caianiello", University of Salerno, Fisciano (SA), Italy}
\affiliation{CNR SPIN Salerno, via Giovanni Paolo II, 132, 84084 Fisciano (SA), Italy}
\author{P.~Sabatino}
\affiliation{Physics Department ``E. R. Caianiello", University of Salerno, Fisciano (SA), Italy}
\affiliation{CNR SPIN Salerno, via Giovanni Paolo II, 132, 84084 Fisciano (SA), Italy}
\author{G.~Carapella}
\affiliation{Physics Department ``E. R. Caianiello", University of Salerno, Fisciano (SA), Italy}
\affiliation{CNR SPIN Salerno, via Giovanni Paolo II, 132, 84084 Fisciano (SA), Italy}
\author{A.~Nigro}
\affiliation{Physics Department ``E. R. Caianiello", University of Salerno, Fisciano (SA), Italy}
\affiliation{CNR SPIN Salerno, via Giovanni Paolo II, 132, 84084 Fisciano (SA), Italy}
\author{S.~Pace}
\affiliation{Physics Department ``E. R. Caianiello", University of Salerno, Fisciano (SA), Italy}
\affiliation{CNR SPIN Salerno, via Giovanni Paolo II, 132, 84084 Fisciano (SA), Italy}
\author{V.~V.~Moshchalkov}
\affiliation{INPAC and Nanoscale Superconductivity and Magnetism Group, K. U. Leuven, Celestijnenlaan 200D, B-3001 Leuven, Belgium}
\author{A.~V.~Silhanek}
\affiliation{D\'epartment de Physique, Universit\'e de Li\`ege, B-4000 Sart Tilman, Belgium}

\date{\today}

\begin{abstract}
We study the instability of the superconducting state in a mesoscopic geometry for the low pinning material $\textrm{Mo}_3$Ge characterized by a large Ginzburg-Landau parameter. We observe that in the current driven switching to the normal state from a nonlinear region of the Abrikosov flux flow, the mean critical vortex velocity reaches a limiting maximum velocity as a function of the applied magnetic field. Based on time dependent Ginzburg-Landau simulations we argue that the observed behavior is due to the high velocity vortex dynamics confined on a mesoscopic scale. We build up a general phase diagram which includes all possible dynamic configurations of Abrikosov lattice in a mesoscopic superconductor.
\end{abstract}

\maketitle

\section{Introduction}
Continuos advancements in nanofabrication have permitted exploration and discovery new emergent physical phenomena when approaching the meso- and nanoscopic limit.\cite{Refv6-1, Ref1, Ref2, Ref3} The stability of the superconducting state under geometrical confinement is now intensively investigated due in part to anomalous mixed state in type-II superconductors,\cite{Ref4, Ref5} as well as for reaching a better performance in their potential applications.\cite{Ref6} Unfortunately, one rarely witnesses the persistence of the non-dissipative regime up to the depairing current density $J_{dp}$, due to the current induced motion of magnetic flux quantum units (Abrikosov vortices) and the consequent Joule heating for currents above a critical current $J_c < J_{dp}$. The actual discrepancy between the theoretical expectation and the experimental fact has its origin in the largely neglected non-equilibrium phenomena occurring at the core of swiftly moving Abrikosov vortices.\cite{Ref10, RefA1} One of these effects, taking place at intermediate current densities $J_c < J^* < J_{dp}$, consists of a deformation of the flux quanta core due to the slow healing time of the superconducting condensate after the passage of a vortex singularity in the condensate.\cite{Ref7} As a consequence, a rapid moving vortex leaves behind a trail of depleted order parameter which further facilitates the motion of other vortices thus forming rivers of flux leading to a net increase of dissipation and triggering an abrupt transition from the Abrikosov flux flow regime to the normal state.\cite{Ref25, RefN11} No matter which mechanism is responsible for this current instability,\cite{Ref7, Ref25, RefN11, Ref8, Ref9, Refv1-2} in this work we show that in the mesoscopic regime, the average critical velocity needed to trigger the instabilities is limited by a maximum speed value, that is not observed under no confinement. We investigate the so far totally unexplored mesoscopic regime where the scenario based on pinning disorder is of no application due to the fact that we use an extremely weak pinning superconducting material, $\textrm{Mo}_3$Ge, in which free flux flow has been recently confirmed.\cite{Ref15} Many studies have been carried out in order to address other possible competing effects such as the influence of pinning properties of the intrinsic material \cite{Ref12, Ref13} and the artificially structured superconductors.\cite{Ref14} Neverthelss, in all these cases the geometry of the test sample has been kept on a macroscopic scale. Here we demonstrate that the Abrikosov lattice instability is affected by a significant surface barrier in mesoscopic superconductors. Time-dependent Ginzburg-Landau simulations satisfactory reproduce the experimental results and give a complete overall description of Abrikosov vortex dynamics driven at high velocity. This can be a general example for high velocity dynamics in any different context of confined geometry, for example in the dynamics of magnetic entities such as skyrmions,\cite{Ref30} in flows and mixing in microfluidic devices,\cite{Ref32} as well as in high speed impact of fluid within a granular material.\cite{Ref33}

\section{Experimental results}
\subsection{Vortex pinning properties}
\begin{figure}
\includegraphics[width=1\columnwidth]{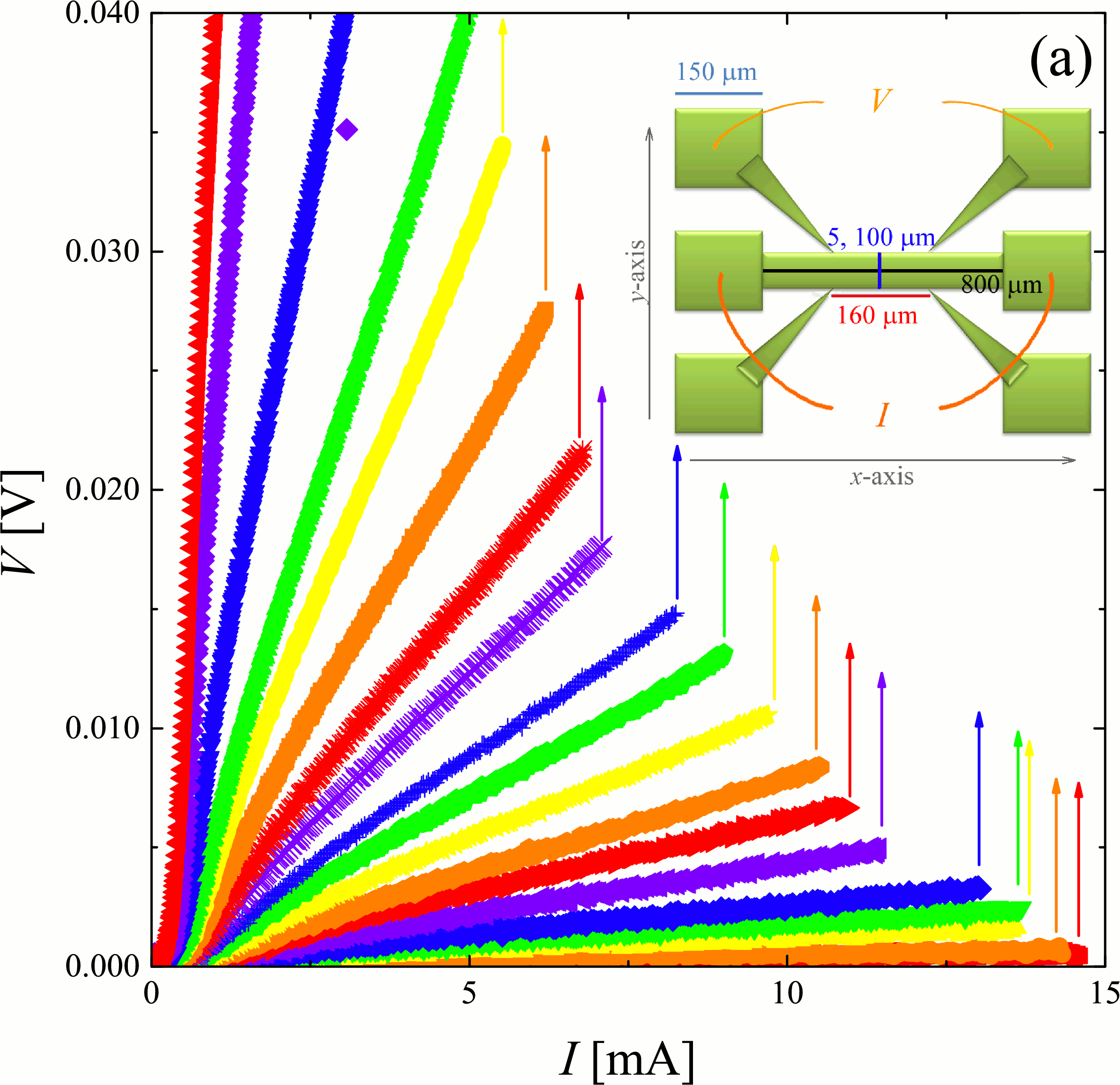}
\\
\includegraphics[width=1\columnwidth]{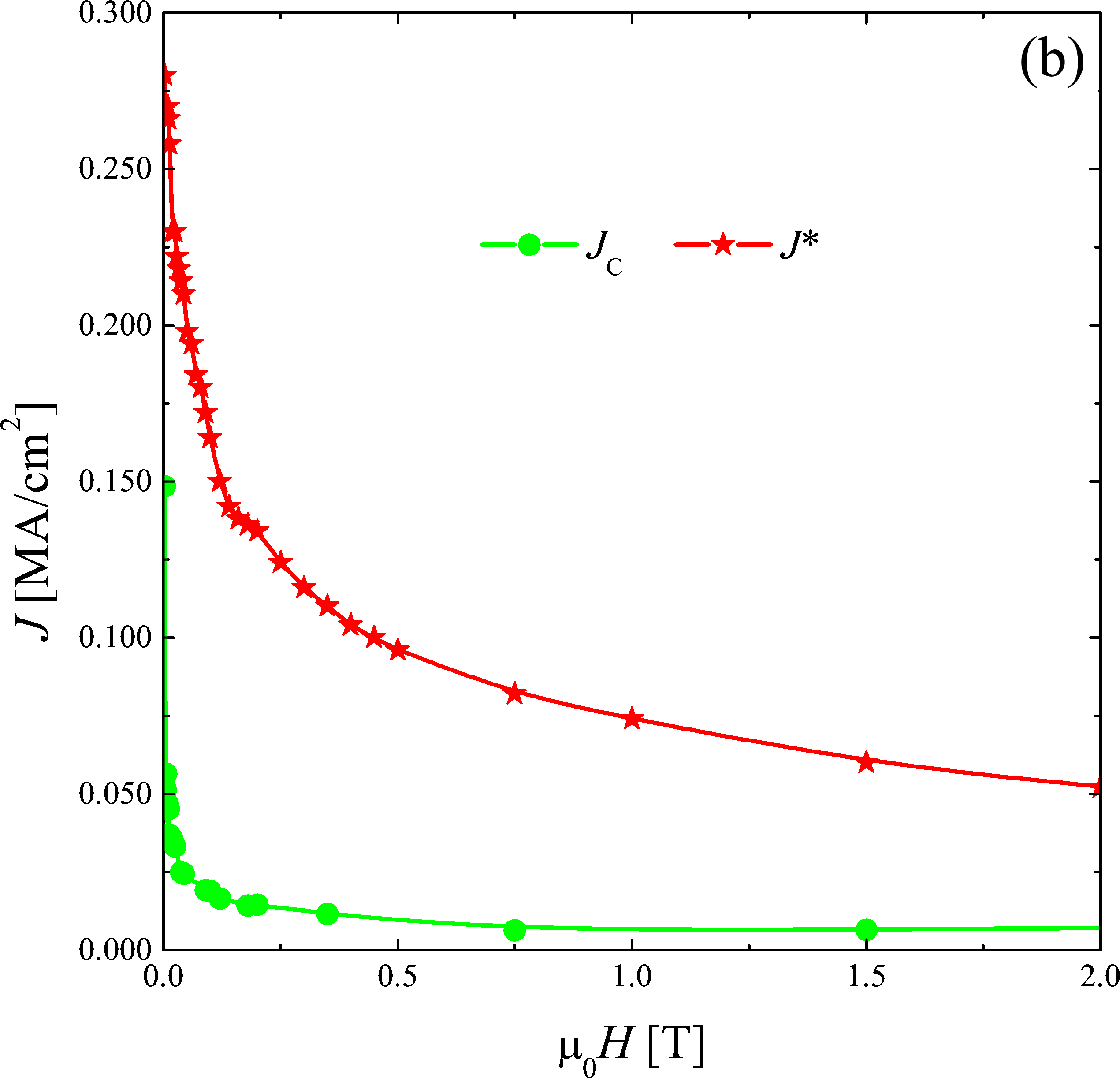}
\caption{\label{Fig1}[color online] (a) The $I$-$V$ curves for the 100~$\mu$m wide strip S100 at $T = 1.6$~K and $H = $ 0.9, 2.5, 6.0, 9.5, 13, 24, 33, 43, 60, 81, 100, 140, 180, 250, 350, 451, 750, 1497, 2494~mT. In the inset the sample layout is sketched. (b) Critical and instability current densities as a function of the applied magnetic field extracted from the $I$-$V$ curves.}.
\end{figure}
$\textrm{Mo}_3$Ge thin films were grown on Si/Si$\textrm{O}_2$ substrates by pulsed laser deposition technique using a Nd:YAG ($\lambda = 532$~nm) pulsed laser of 55~J energy and a repetition rate of 10 Hz. The deposition was performed at a pressure of $10^{-7}$~mbar. Using these parameters a deposition rate of 1.1~nm/min is achieved. All depositions were done on Si wafers with an amorphous SiO$_x$ top-layer. Microbridges were obtained by electron beam lithography with thickness $d = 50$~nm, length $L = 160$~$\mu$m and different linewidths $w = 5 \div 100$~$\mu$m.\cite{RefSM1}

The superconducting properties strongly depend on the thickness of the films. For 4~nm thick films, there is no superconducting transition. For films thicker than 25~nm the superconducting transition saturates reaching values up to 7~K.\cite{RefSM1} Typical values of the superconducting parameters for the highest $T_c$ film, namely S100, are: $T_c = 6.5$~K, $H_{c2}(0) \sim 9$~T, $J_c(0) = 0.15$~MA/cm$^2$. In addition it has been shown that usually the irreversible magnetization loops are already closed at 20~mT, i.e. two orders of magnitude smaller than $H_{c2}$.\cite{RefSM2}

Pulsed current-voltage ($I$-$V$) measurements were perfomed in order to minimize self-heating effects.\cite{Ref12} Since unavoidable self-heating may affect experimental data, first of all we chose a pulsed biasing mode with a pulse width of 2.5~ms and an inter-pulse period of 1~s.\cite{RefSM3}

In Fig. 1 the current-voltage ($I$-$V$) curves for the measured S100 macroscopic strip are shown along with the data related to the critical current density $J_c$ and the instability current $J^*$ as a function of the applied magnetic field. We note that for the whole field range, $J_c$ is considerably smaller than $J^*$, and exhibits a very steep decrease as a function of the magnetic field. The absolute $J_c$ values of the order of $10^4$~A/cm$^2$ reveal the weak pinning nature of this material.

\subsection{Vortex lattice instability at the mesoscopic scale}
The mesoscopic limit is reached when $d \ll \lambda$,\cite{Ref19, Ref20, Ref21} with $\lambda$ the London penetration depth, and the sample width is narrower than the Pearl length $w \leq \Lambda = 2\lambda^2/d$,\cite{Ref19} and much wider than the Ginzburg-Landau coherence length $w \gg \xi_{GL}$. The estimated values of effective coherence length and penetration depth at very low temperatures for our $\textrm{Mo}_3$Ge films are as low as $\xi_{GL} \simeq 5$~nm and as large as $\lambda = 500$~nm, corresponding to a large Ginzburg-Landau parameter $k = 100$. The conditions for the mesoscopic limit are reasonably satisfied for our thin films of $w = 5$~$\mu$m being $\Lambda = 10$~$\mu$m. On these mesoscopic samples we checked that self-heating effects can be neglected. Indeed, no hysteresis occurs in the $I$-$V$ curves by performing measurements forth and back (increasing and subsequently decreasing current) of each curve by current biasing. Moreover, we can assure that the instability point of each curve remains unchanged and it is always reproducible, although at low fields the metastable states can change before the normal state is reached. We also took into account self-heating by considering the Bezuglyi-Shklovskij approach for the term of quasi-particle overheating,\cite{Ref8} leading to the estimate of the threshold magnetic field value $B_T = 0.374 e h \tau_E / k_B \sigma_N d \sim 3$~T, where $h$ is the heat transfer coefficient to the coolant,\cite{RefSM4+} $\tau_E$ is the quasi-particle relaxation time,\cite{Ref15} $\sigma_N$ is the normal conductivity.\cite{RefSM1} In other words, heating effects become significant for $B > B_T$, out of the field range in which the maximum speed limit of the moving Abrikosov lattice is achieved. In addition, we derived from the $I$-$V$ data of Fig. 1a the dissipated power $P^* = I^* V^*$, which is an increasing function of the magnetic field, as shown in Fig. 2. This is the experimental evidence that thermal effects are not determining the flux flow instability points. Indeed, if this was the case of a thermal runway, $P^*$ should be independent of magnetic field, as already pointed out by Xiao \emph{et al.}.\cite{RefSM5} We also note that this magnetic field dependence has been predicted by Vina \emph{et al.}\cite{RefSM6} on self-heating based calculations. However they deal with an high temperature superconductor, YBCO microbridges, in a temperature range close to $T_c$, $0.8 < T/T_c < 1$, whose normal state resistivity are several orders of magnitude larger than in our mesoscopic strips.
\begin{figure}
\includegraphics[width=1\columnwidth]{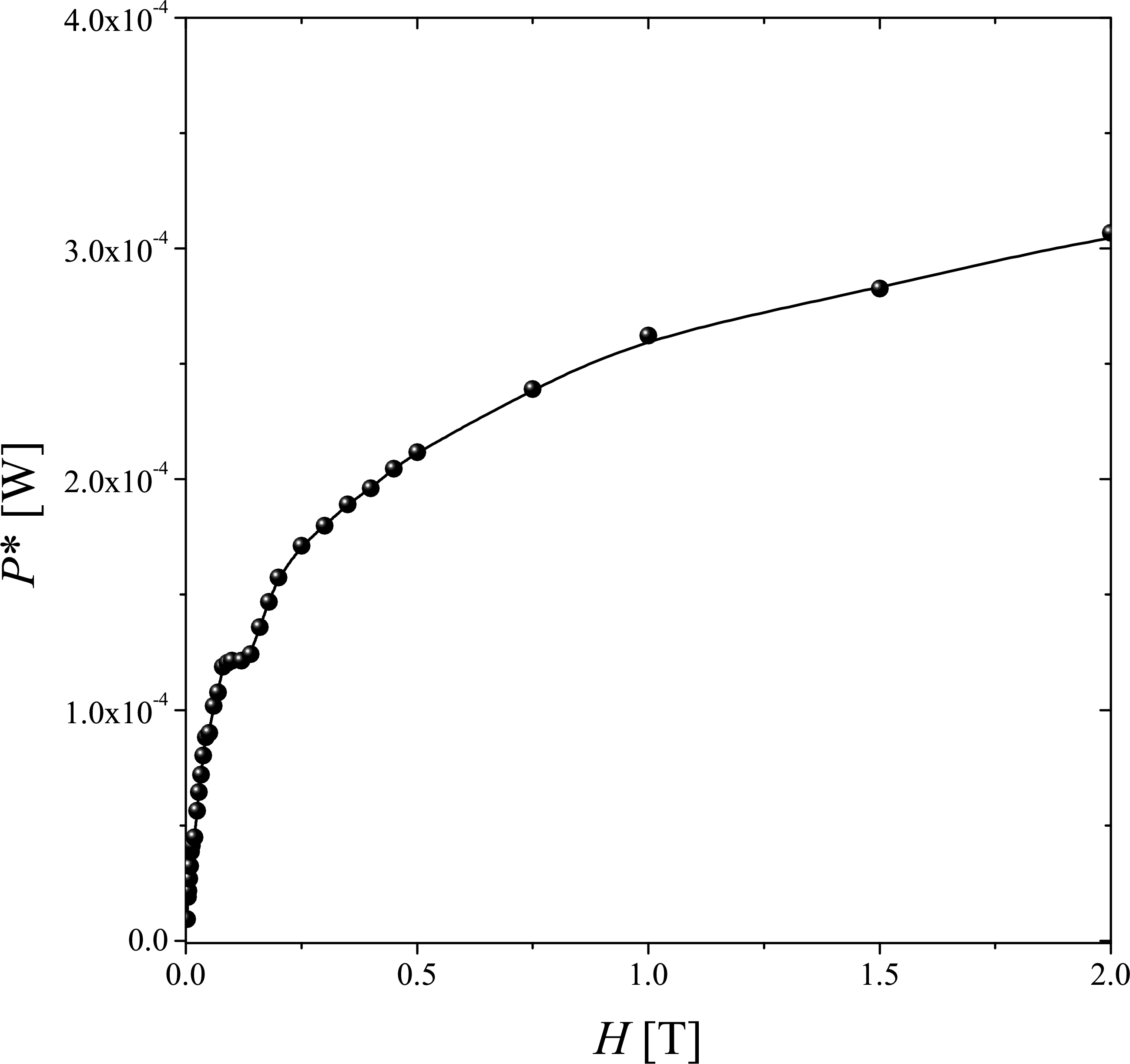}
\caption{\label{Fig2}The dissipated power as a function of the magnetic field. Each data is estimated at the instability points marked by the arrows in Fig. 1(a).}
\end{figure}

In Fig. 3(a) we report the $I$-$V$ curves as a function of magnetic field for the 5~$\mu$m mesoscopic strip Sx at the lowest temperature $T = 1.6$~K. The inset shows a single curve measured at low field, in which multiple voltage jumps are observed in the $V(I)$ branch above the instability point $(I^*, V^*)$ and up to the normal current $I^*_N$. In this case, the transition to the normal state follows several metastable states, instead of being an abrupt voltage jump.

\begin{figure}[t]
\includegraphics[width=1\columnwidth]{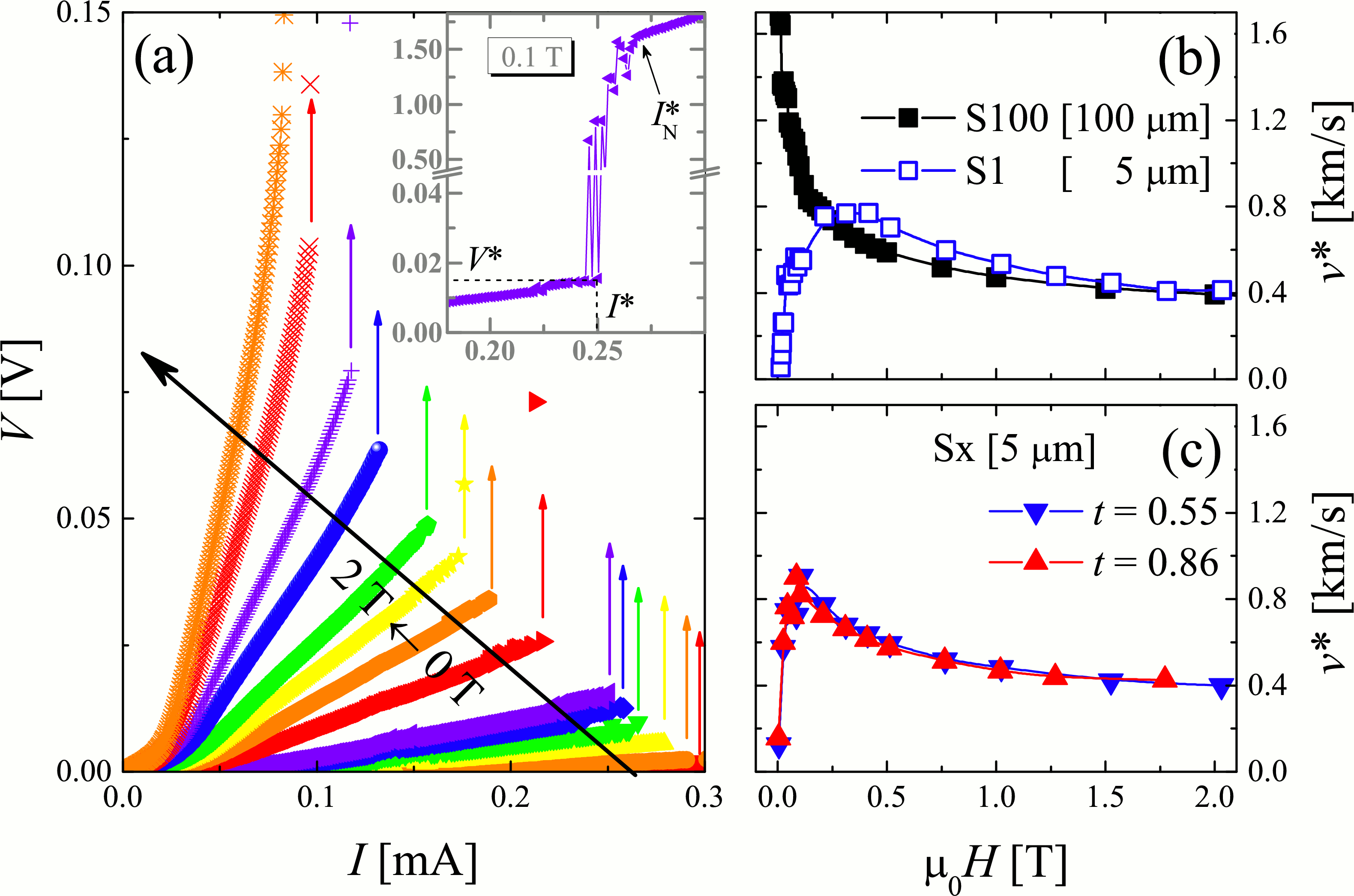}
\caption{\label{Fig3}[color online] (a) Experimental $I$-$V$ curves at $T = 1.6$~K for increasing magnetic field (as indicated by the arrow, in mT): 4, 25, 46, 67, 87, 109, 209, 311, 412, 513, 767, 1023, 1526, 2031. The inset shows several multiple jumps before reaching the normal resistance. (b) The critical vortex velocity as a function of magnetic field for the S100 and S1 strips, indicated by the full and open symbols, respectively. (c) The $v^*(\mu_0 H)$ curves for two reduced temperatures $t = 0.55$ and $0.86$ measured on the Sx strip. Lines are guide to the eye.}
\end{figure}

Being $\textrm{Mo}_3$Ge considered a weak pinning superconductor, almost linear flux flow branches are also expected up to high bias current, as it is noticed both for Sx (see Fig. 3(a)) and the 100~$\mu$m strip S100 (see Fig. 1(a)). From the last point $(I^*, V^*)$ marked by the arrows in the continuous branch, we estimate the mean critical velocity of the moving vortex lattice as $v^* = E^*/\mu_0 H$. By extracting the $v^*(\mu_0 H)$, we obtain a surprising result: the size reduction down to the mesoscopic scale implies the change of the critical velocity behavior in a substantial magnetic field range $\mu_0 H < 0.5$~T, as shown in Fig. 3(b) for the case of the 5~$\mu$m mesoscopic strip S1 and the S100 macroscopic one. Interestingly, from the fact that $v^*$ estimations acquired at two different temperatures (see Fig. 3(c)) show no difference, we can suggest that the observed change in $v^*$ for the mesoscopic sample is rather $T$ idependent.

\subsection{Pinning effect on vortex instability}
In order to investigate if the observed behavior $v^*(\mu_0H)$ in the mesoscopic limit is influenced by any bulk pinning, we changed the intrinsic pinning from the weak $\textrm{Mo}_3$Ge thin films to a well known stronger pinning superconductor, namely NbN.

We fabricated a mesoscopic strip of 1~$\mu$m width and 10~$\mu$m length, realized by e-beam litography on $d = 20$~nm thin film, so that we obtained $\lambda(0) = 400$~nm and $\Lambda = 16$~$\mu$m. Carrying out the same data analysis performed in the case of the $\textrm{Mo}_3$Ge samples, we obtained the results collected in Fig. 4, where the critical voltages of the NbN sample are plotted together with the data related to the $\textrm{Mo}_3$Ge sample S1. We find that even a strong pinning material on the mesoscopic scale has the same striking behavior, although on a larger magnetic field range.
\begin{figure}[t]
\includegraphics[width=1\columnwidth]{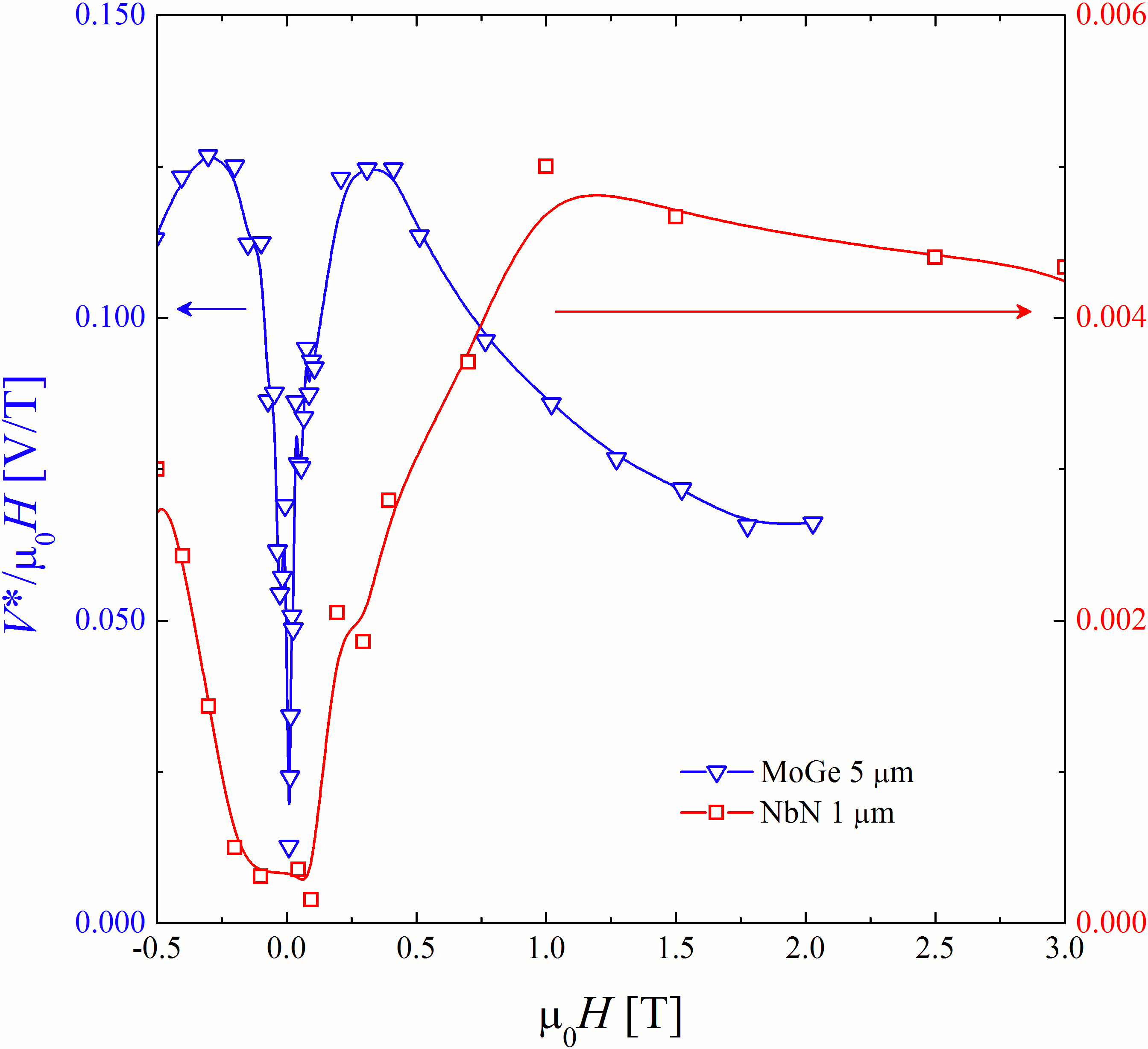}
\caption{\label{Fig4}[color online] Critical voltages vs magnetic field. Data acquired on a mesoscopic NbN stronger pinning superconducting strip (squares) compared with the $\textrm{Mo}_3$Ge S1 mesoscopic strip (triangles).}
\end{figure}

\section{Numerical simulations}
\subsection{Model}
Time-dependent Ginzburg-Landau (TDGL) simulations are used to gain information on the vortex dynamics accounting for the $I$-$V$ curves observed in the $\textrm{Mo}_3$Ge mesoscopic superconductor. We use the TDGL model in its 2D simplified form, being justified when the strip exhibits a large parameter $k$ and it is in the mesoscopic limit, that is our experimental case. In the mesoscopic limit current density is reasonably uniform and the magnetic field which is induced by the transport and screening currents can be usually neglected.\cite{Ref19, Ref20, Ref21} Moreover, in the model we neglect intrinsic bulk pinning, according to experimental data.

In the numerical simulations we assume the strip width equal to $w = 160 \xi_L$ in the $x$-direction and length $L = 80 \xi_L$ in the $y$-direction. Experimentally, the phenomenon we want to describe is found almost independent of temperature, $v^*(\mu_0 H,T)$, and it is recorded up to the reduced temperature $t = 0.86$. At this reduced temperature the normalized width of the real strips is $w = 370 \xi_L$. Indeed, we use the smaller width $w = 160 \xi_L$ in the simulations either because we checked that vortex dynamics involved did not change appreciably if such dimensions were further increased and because would be more cumbersome to present snapshots of vortex dynamics covering the larger width of $370 \xi_L$, being vortex cores extended for only about $3 \xi_L$.

We also underline that in the non-mesoscopic regime with no bulk pinning (see, e.g., Ref.\onlinecite{Ref25}) the current distribution is strongly peaked at edges of the strip and the current-assisted vortex nucleation is present also at fields $H < H_S/2$, where $H_S$ is the field at which static vortices are present in the strip also for $J = 0$.\cite{Ref20, Ref21, Ref24} In particular, at zero applied field vortices and antivortices nucleate at edges and annihilate at center of the strip.\cite{Ref25} On the contrary, the region $0 < H < H_S/2$ of hampered vortex nucleation is observed in a mesoscopic system compelling a vortex velocity increase with magnetic field from zero to a maximum value. This region can be hidden in the macroscopic system, where the finite (maximum) value of the average velocity can be achieved already at very low fields close to $H = 0$. This may account for the different behavior of $v^*(\mu_0 H)$ in the mesoscopic strip with respect to the macroscopic case.

In the following we will assume to work in a temperature range so close to $T_c$, that the phenomenological TDGL model is supposedly adequate. The 2D TDGL equation for the complex order parameter $\psi = |\psi|e^{i\phi}$ takes the form:\cite{Ref21, Ref22, Ref23}
\begin{equation}
u \left( \frac{\partial}{\partial t} + i \phi \right) \psi = \left( \nabla - i \textbf{A} \right)^2 \psi + \left( 1 - |\psi|^2 \right) \psi
\end{equation}
coupled with the equation for the electrostatic potential $\nabla^2 \phi = \textrm{div}\left[\text{Im}\left(\psi^*\left(\nabla-i\textbf{A}\right)\psi\right)\right]$, where \textbf{A} is the vector potential associated to the external magnetic field $H$, $\phi$ is the electrostatic potential and the coefficient $u = 5.79$ governs the relaxation of the order parameter.\cite{Ref26} All physical quantities are measured in dimensionless units:\cite{Ref21, Ref22, Ref23} the coordinates are in units of the coherence length $\xi_{GL}(T)$, time is in units of the relaxation time $\tau$, the order parameter is in units of the superconducting gap $\Delta(T)$, the vector potential is in units of $\Phi_0 / 2 \pi \xi_{GL}$ ($\Phi_0$ is the quantum of magnetic flux), and the electrostatic potential is in units of $\phi_0(T) = \hbar / 2 e \tau$. In these units the magnetic field is scaled with $H_{c2}(T) = \Phi_0 / 2 \pi \xi_{GL}^2$ and the current density with $j_0(T) = c \Phi_0 / 8 \pi^2 \lambda^2 \xi_{GL}$.  The field $H$ is applied in the $z$ direction and the current density $J$ is applied in the y-direction. We make use of the ``bridge'' boundary condition in the $y$-direction and of an insulator-superconductor boundary condition in the $x$-direction.\cite{Ref21, Ref22, Ref23}

\subsection{Flux flow results}
\begin{figure}[t]
\includegraphics[width=1\columnwidth]{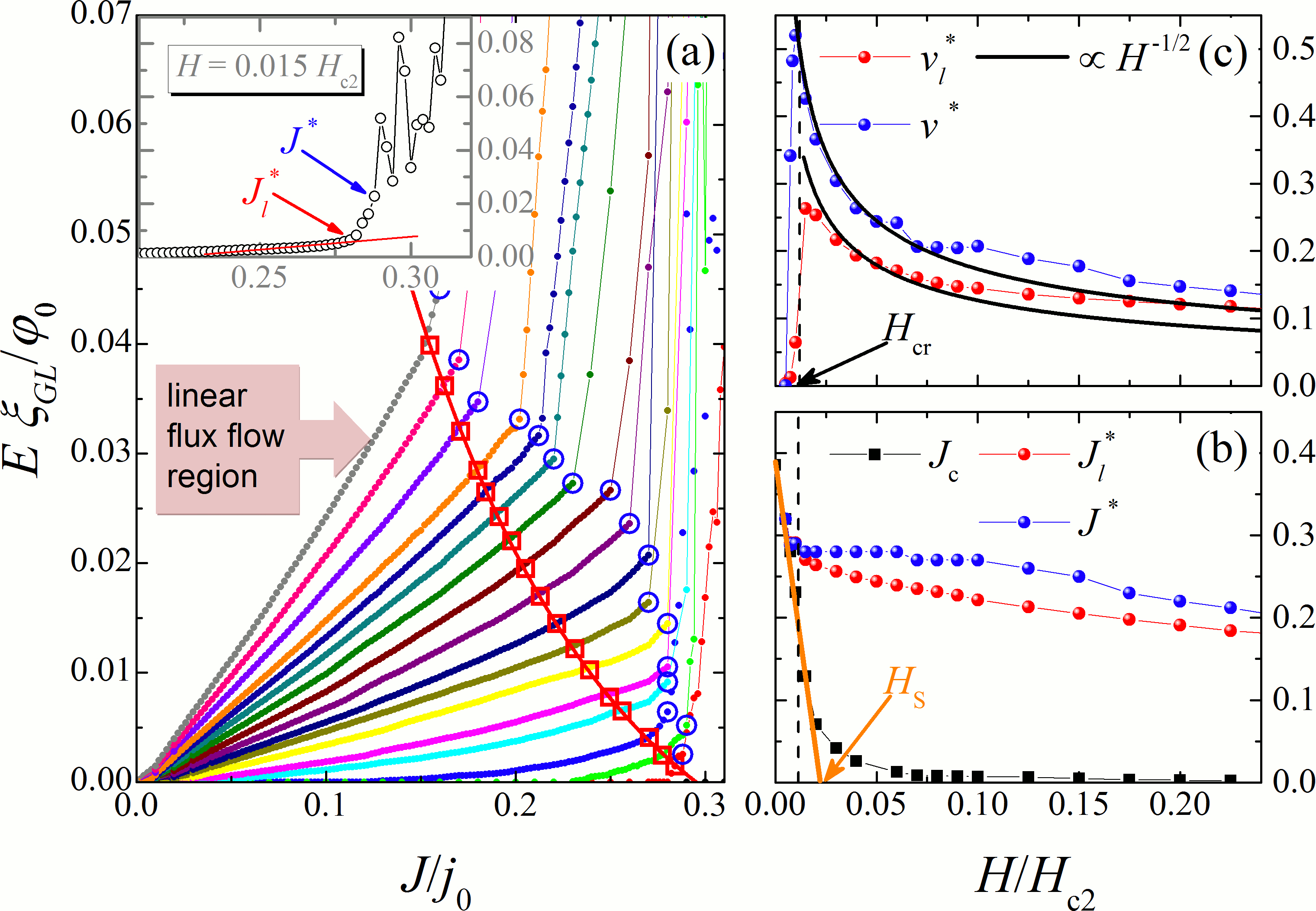}
\caption{\label{Fig5}[color online] Numerical results: (a) $E(J)$ curves for different magnetic field values in the low electric field range. In units of $H_{c2}$ the field
values are 0.0075 (red dots), 0.01, 0.015, 0.03, 0.04, 0.06, 0.08, 0.1, 0.125, 0.15, 0.175, 0.2, 0.225, 0.25, 0.3,  0.35, 0.4 (grey dots).
The inset shows a single curve at low field. (b) Critical and instability current densities as a function of magnetic field. (c) Average vortex critical velocities for the linear and nonlinear regimes as a function of magnetic field.}
\end{figure}
\begin{figure*}[t]
\includegraphics[width=1\textwidth]{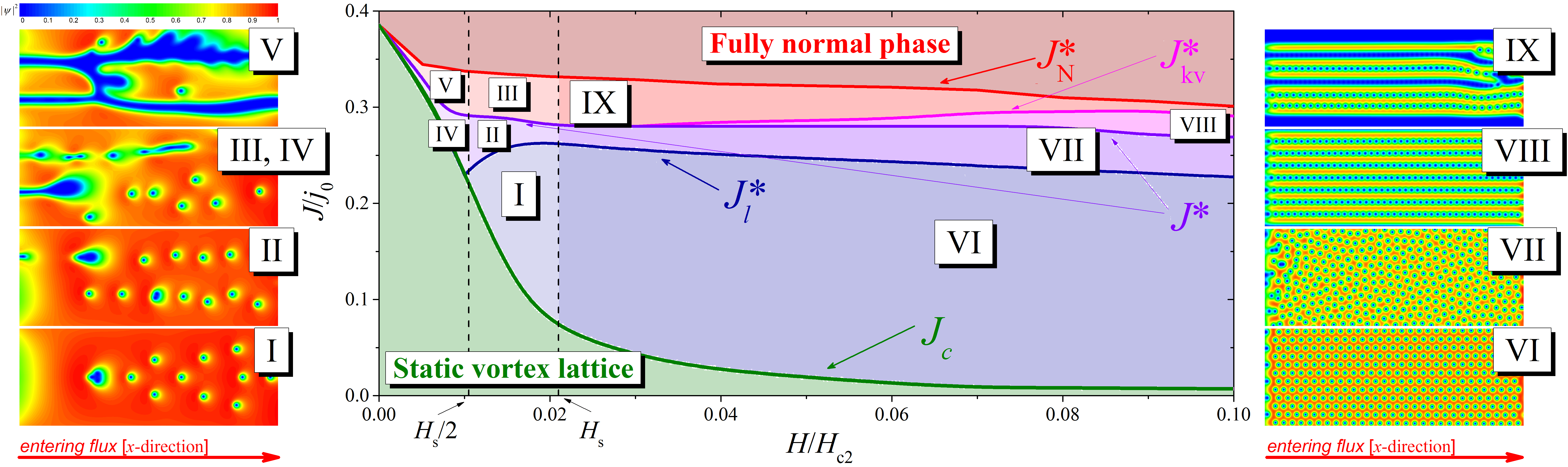}
\caption{\label{Fig6}[color online] Numerically obtained dynamic phase diagram of vortex lattice in a mesoscopic superconductor. It is identified the peculiar ``entry field'' $H = H_S$ above which a regular vortex lattice can be set in motion, as well as the instability current density $J^*$ above which moving ordinary vortices transform in kinematic vortices at moderate/high fields or ordinary vortex bundles mixed to clusterized normal regions at low field.  When present, normal regions expand at expense of vortex matter with increasing current up to a final $J^*_N$ where the system undergoes the transition to the fully normal state. Snapshots show the evolution of the vortex lattice configuration on increasing the driving current.}
\end{figure*}
In Fig. 5(a), we show the calculated $E(J)$ curves, in the low electric field range, for several values of magnetic field $H$ applied perpendicular to the strip. Curves display a fully linear (at moderate and high fields) or nearly linear (at low fields) flux-flow branch starting at some critical current $J_c$ in the rather large current range $J_c < J < J_l^*$, followed by a deviation from linear behavior in the limited current range $J_l^* < J < J^*$ and ending with a more or less abrupt transition to the fully normal state. The red line is displayed as a guide to the eye to mark the critical points $(E_l^*, J_l^*)$ where the departure from almost linear behavior occurs. The circles mark the instability points $(E^*, J^*)$ at which the continuos nonlinear branch ends and a very high differential resistivity branch (at moderate/high fields) or a meta-stable branch (at low fields) is followed before transition to the normal state is achieved. The Fig. 5(b) shows the relevant current densities $J_l^*$, $J^*$, $J_c$ as a function of magnetic field. The average critical velocity of the linear $v_l^* = E_l^*/\mu_o H$ and nonlinear $v^* = E^*/\mu_o H$ regimes are plotted as a function of applied magnetic field in Fig. 5(c). In analogy with experimental data, both critical velocities exhibit a non-monotonic behavior with a maximum at a certain field $H_{cr}$, after which a decreasing function of $H$ is established, approximately as $H^{-1/2}$.\cite{Ref12} In the following we focus on the vortex dynamics accounting for the $E(J)$ in the magnetic field range around $H_{cr}$ where such a crossover is found (see Fig. 5(c)) and $v^*_{max}$ is reached. In Fig. 5(b) we can distinguish, in the magnetic field range in which the critical velocity is increasing, the ``entry field'' $H = H_S$, that is $H_S = 0.021 H_{c2}$. Inspection of Fig. 5(b) and 5(c) suggests that the crossover field $H_{cr}$ can essentially be identified with $H_s/2$.

\subsection{Nonequilibrium phase diagram}
At magnetic fields larger than the entry field $H_S$ a regular Abrikosov vortex lattice is expected to be present even at $J = 0$. At fields $H_S/2 < H < H_S$, by increasing the driving current, vortices nucleate to the left edge of the strip but start to flow only at some finite current $J_c$, due to the presence of a surface barrier in the system.\cite{Ref20, Ref21} Though a triangular vortex lattice is not fully created, there exists a quasi-ordered motion of vortices [see snapshot I in Fig. 6], that results in an almost linear branch in the $E(J)$ curve up to $J_l^*$. By further increasing current, a departure from nearly linear $E(J)$ curve occurs up to the critical current $J^*$, and vortex flow transforms to a row-like structure, as visualized in snapshot II. At currents larger than $J^*$ the row structure evolves into normal channel-like configuration, as shown in snapshot III. This corresponds to a noisy behavior in the $E(J)$ curve and to dynamical states in which the presence of vortices is restored at the expense of normal channels (see inset of Fig. 5(a)). For fields $H < H_S/2$, we note only the sequence of states which in our analysis involves the transition to the normal state with a more or less pronounced intermittent effect [see snapshot IV and V]. Interestingly, this intermittence is also observed in the experimental curves [see inset of Fig. 3(a)]. We should remark that, for the magnetic field range $H_S/2 < H <H_S$, there exists a finite current range where nearly ordered vortex matter is driven by the bias current. In this lower field region vortices can nucleate in the strip only when a quite large uniformly distributed transport current ($\sim J_{dp}$) helps the screening current to suppress the order parameter at one of the edges of the strip, thus promoting vortex nucleation.\cite{Ref21, Ref24, RefAVS1}

In Fig. 6 we include the full zoology of vortex lattice phases in motion, with particular attention on the dynamic phase diagram in which non-linearity arises. Here some well-known phases are reported for completeness: the static vortex lattice which exists only for $J < J_c$ and the fully normal phase for $J > J^*_N$. In the low field region $H_S/2<H<H_S$, Region I corresponds to the current assisted vortex nucleation with an almost triangular moving vortex lattice, that results in the almost linear flux flow motion $J_c < J < J_l^*$. At larger currents we find other two possible nonequilibrium phases, II and III. In Region II ($J_l^* < J < J^*$) a row-like moving vortex lattice can exist with a corresponding filamentary displacement of vortices (vortex river). For larger currents clusters of normal metal phase develop which match the intermittence phase III with metastable states before reaching the fully normal phase. In agreement with analytical predictions presented in Ref. \onlinecite{Ref24}, in the field range $0 < H < H_S/2$ transport-current-assisted discrete vortex nucleation is practically absent. Here, we only observe a disordered bundlelike vortex nucleation and motion  accounting for a strongly nonlinear $E(J)$ branch up to $J^*$ (Region IV), followed by a regime (Region V) where flux bundles are mixed to a clusterized normal phase, as shown in snapshot V of Fig. 6. At $H > H_S$, Region VI corresponds to the ordinary linear flux flow accounted by a moving triangular lattice (snapshot VI). This regular motion is observed up to a critical current $J_l^*$ resulting in a nearly linear branch of the $E(J)$ curve. Region VII marks the nonlinear flux flow motion for $J_l^*< J < J^*$ accounted for a moving glassy lattice (snapshot VII). Above the instability current $J^*$, jumps to high resistivity branches can occur (Region VIII). These branches, which extend in the range $J^*< J < J_{kv}^*$, are accounted for a channel-like structure of vortices as shown in snapshot VIII, very similar to the one described in Ref. \onlinecite{Ref25}. For $J > J_{kv}^*$, the vortex channel structure leads to the opening of normal channels (see snapshot IX), which are responsible for a more less abrupt transition to the highest resistive state, i.e. the normal state. Fast (kinematics) vortices surfing on channels of very depressed superconductivity shown in snapshot VIII were investigated in detail in Ref. \onlinecite{Ref25} within the generalized TDGL model that accounts for the non-equilibrium effects through a parameter $\gamma$.\cite{Ref26} In our simulations we used the standard TDGL with $\gamma = 0$ and, though present, the high resistivity branches accounted for kinematic vortices are consistently \cite{Ref26} recovered only in a current range much narrower than the one found in Ref. \onlinecite{Ref25}.

\section{Conclusions}
In the framework of vortex dynamics we identify the limiting behavior of the Abrikosov lattice stability driven at high vortex velocity in the absence of bulk pinning, with the only constrain of confined mesoscopic geometry. The possibility to reach a maximum critical velocity as a function of the applied magnetic field is naturally explained by the TDGL phenomenological approach, which allows us to give a complete view of nonequilibrium vortex phases tunable by the external magnetic field and/or the bias current in a dynamic phase diagram. In a mesoscopic superconductor the physical meaning of such speed limit to the Abrikosov vortex velocity is strictly connected to the presence of a surface pinning, namely an edge barrier, which first hampers and then delays the lattice motion, due to the rearrangement of vortices configuration in rows-like flow rather than keeping the usual ordered triangular vortex lattice in motion. To visualize it, real-space images of the driven lattice show that motion occurs along channels that are aligned with the direction of the driving force and periodically spaced in the transverse direction. Phase slips, however, occur at the channel boundaries, indicating that channels become uncorrelated at very high driving current, before the instability takes place. In our case those channels may exist only at a velocity $v < v^*(H_{cr}) = v^*_{max}$, that is the speed limit for the moving Abrikosov lattice.

The sequence in which these dynamical phases appear at high bias currents is usually nontrivial, and the simplified models of vortices as pointlike classical particles seem to have missed what a more realistic approach based on TDGL formalism is able to catch.\cite{Ref29}

A further comparison of data on weak $\textrm{Mo}_3$Ge superconductor with the bulk pinning effects in NbN surprisingly led to similar results, thus conferring to our findings even more generality. Our results demonstrate that geometric reduction on mesoscopic scale can radically change the dissipative regimes in superconducting materials, thus improving the performance of those devices based on superconducting nanostructures.

\begin{acknowledgments}
We thank L. Baez Alonso for the fabrication of $\textrm{Mo}_3$Ge thin films and M.V. Milosevic for fruitful discussions. A.L. acknowledges financial support from PON Ricerca e Competitivit\`a 2007-2013 under Grant Agreement PON NAFASSY, PONa3\_00007.
\end{acknowledgments}


\begin{thebibliography}{50}

\bibitem{Refv6-1} M.P. Das, B.J. Wilson, Adv. Sci. Mat.: Nanosci. Nanotechnol. \textbf{6}, 013001 (2015).
\bibitem{Ref1} S. Nawaz, R. Arpaia, F. Lombardi, T. Bauch, Phys. Rev. Lett. \textbf{110}, 167004 (2013).
\bibitem{Ref2} K. Xu, P. Cao, J.R. Heath, Nano Lett. \textbf{10}, 4206 (2010).
\bibitem{Ref3} R. Cordoba \emph{et al.}, Nat. Commun. \textbf{4}, 1437 (2013).
\bibitem{Ref4} S.G. Doettinger, R.P. Huebener, R. Gerdemann, A. Kuhle, S. Anders, T.G. Trauble, J.C. Villegier, Phys. Rev. Lett. \textbf{73}, 1691 (1994).
\bibitem{Ref5} Yu Chen, Yen-Hsiang Lin, S.D. Snyder, A.M. Goldman, A. Kamenev, Nat. Phys. \textbf(10), 567 (2014); I. Lukyanchuk \emph{et al.}, Nat. Phys. \textbf{11}, 21 (2015).
\bibitem{Ref6} M.N. Kunchur, D.K. Christen, C.E. Klabunde, K. Salama, Appl. Phys. Lett. \textbf{67}, 848 (1995).
\bibitem{Ref10} A.G. Sivakov, A.M. Glukhov, A.N. Omelyanchouk, Y. Koval, P. Muller, A.V. Ustinov, Phys. Rev. Lett. \textbf{91}, 267001 (2003).
\bibitem{RefA1} R.S. Keizer, M.G. Flokstra, J. Aarts, T.M. Klapwijk, Phys. Rev. Lett. \textbf{96}, 147002 (2006).
\bibitem{Ref7} A.I. Larkin, Y.N. Ovchinnikov, Nonequilibrium Superconductivity, Elsevier, Amsterdam, p. 493 (1986).
\bibitem{Ref25} D.Y. Vodolazov, F.M. Peeters,  Phys. Rev. B \textbf{76}, 014521 (2007).
\bibitem{RefN11} G.R. Berdiyorov, M.V. Milo\v sevi\'c, F. M. Peeters, Phys. Rev. B \textbf{79}, 184506 (2009); A.V. Silhanek \emph{et al.}, Phys. Rev. Lett. \textbf{104}, 017001 (2010).
\bibitem{Ref8} A. Bezuglyj, V. Shklovskij, Physica C \textbf{202}, 234 (1992); J. Maza, G. Ferro, J.A. Veira, F. Vidal, Phys. Rev. B \textbf{78}, 094512 (2008).
\bibitem{Ref9} M.N. Kunchur, Phys. Rev. Lett. \textbf{89}, 137005 (2002).
\bibitem{Refv1-2} I. Aranson, B.Ya. Shapiro, V. Vinokur, Phys. Rev. Lett. \textbf{76}, 142 (1996).
\bibitem{Ref15} M. Liang, M.N. Kunchur, J. Hua, Z. Xiao, Phys. Rev. B \textbf{82}, 064502 (2010); M. Liang, M.N. Kunchur, Phys. Rev. B \textbf{82}, 144517 (2010).
\bibitem{Ref12} G. Grimaldi, A. Leo, A. Nigro, S. Pace, R.P. Huebener, Phys. Rev. B \textbf{80}, 144521 (2009).
\bibitem{Ref13} G. Grimaldi, A. Leo, A. Nigro, A.V. Silhanek, N. Verellen, V.V. Moshchalkov, M.V. Milosevic, A. Casaburi, R. Cristiano, S. Pace, Appl. Phys. Lett. \textbf{100}, 202601 (2012).
\bibitem{Ref14} A.V. Silhanek \emph{et al.}, New J. Phys. \textbf{14}, 053006 (2012).
\bibitem{Ref30} S.Z. Lin, C. Reichhardt, C.D. Batista, A. Saxena, Phys. Rev. Lett. \textbf{110}, 207202 (2013).
\bibitem{Ref32} T.V. Nizkaya, E.S. Asmolov, Jiajia Zhou, F. Schmid, O.I. Vinogradova, Phys Rev. E \textbf{91}, 033020 (2015).
\bibitem{Ref33} A.H. Clark, A.J. Petersen, L. Kondic, R.P. Behringer, Phys. Rev. Lett. \textbf{114}, 144502 (2015).
\bibitem{RefSM1} A. Leo, G. Grimaldi, A. Nigro, E. Bruno, F. Priolo, S. Pace, Physica C \textbf{503}, 140 (2014).
\bibitem{RefSM2} M. Motta, F. Colauto, W.A. Ortiz, J. Fritzsche, J. Cuppens, W. Gillijns, V.V. Moshchalkov, T.H. Johansen, A. Sanchez, A.V. Silhanek, Appl. Phys. Lett. \textbf{102}, 212601 (2013).
\bibitem{RefSM3} G. Grimaldi, A. Leo, C. Cirillo, A. Casaburi, R. Cristiano, C. Attanasio, A. Nigro, S. Pace, R. P. Huebener, J. Supercond. Nov. Magn. \textbf{24}, 81 (2011).
\bibitem{Ref19} J.R. Clem, K.K. Berggren, Phys. Rev. B \textbf{84}, 174510 (2011).
\bibitem{Ref20} J.R. Clem, Y. Mawatari, G.R. Berdiyorov, F.M. Peeters, Phys. Rev. B \textbf{85}, 144511 (2012).
\bibitem{Ref21} D.Y. Vodolazov, Phys. Rev. B \textbf{88}, 014525 (2013).
\bibitem{RefSM4+} M.N. Wilson, \emph{Superconducting Magnets} edited by R.G. Scurlock (Oxford University press, Oxford, England, 1983).
\bibitem{RefSM5} Z.L. Xiao, P. Voss-de Haan, G. Jacob, T. Kluge, P. Haibach, H. Adrian, E.Y. Andrei, Phys. Rev. B \textbf{59}, 1481 (1999).
\bibitem{RefSM6} J. Vina, M.T. Gonzalez, M. Ruibal, S.R. Curras, J.A. Veira, J. Maza, F. Vidal, Phys. Rev. B \textbf{68}, 224506 (2003).
\bibitem{Ref24} G.M. Maksimova, N.V. Zhelezina, I.L. Maksimov, Europhys. Lett. \textbf{53}, 639 (2001).
\bibitem{Ref26} L. Kramer, R.J. Watts-Tobin, Phys. Rev. Lett. \textbf{40}, 1041 (1978).
\bibitem{Ref22} P. Sabatino, G. Carapella, G. Costabile, Supercond. Sci. Technol. \textbf{24}, 125007 (2011).
\bibitem{Ref23} G. Carapella, P. Sabatino, G. Costabile, J. Phys.: Condens. Matter \textbf{23}, 435701 (2011).
\bibitem{RefAVS1} D. Cerbu \emph{et al.}, New J. Phys. \textbf{15}, 063022 (2013).
\bibitem{Ref29} O. Iaroshenko, V. Rybalko, V.M. Vinokur, L. Berlyand, Sci. Rep. \textbf{3}, 1758 (2013).

\end{thebibliography}
\end{document}